\documentclass{aastex701}

\shorttitle{EI Cancri with VLITE}
\shortauthors{Silverstein et al.}

\definecolor{pink2}{RGB}{255,0,200}
\definecolor{teal}{RGB}{30,190,180}

\newcommand{\EI}{EI~Cancri }
\newcommand{\EIAB}{EI~Cancri~AB }

\usepackage{multirow}

\begin{document}

\title{First Detection of an Ultracool Dwarf at 340 MHz: VLITE Observations of EI Cancri AB}

\author[0000-0003-2565-7909]{Michele L. Silverstein}
\altaffiliation{NRC Research Associate}
\affiliation{Naval Research Laboratory,
4555 Overlook Avenue SW,
Washington, DC 20375, USA}
\email{mlsilverstein@proton.me}

\author[0000-0001-6812-7938]{Tracy E. Clarke}
\affiliation{Naval Research Laboratory,
4555 Overlook Avenue SW,
Washington, DC 20375, USA}
\email{tracy.e.clarke2.civ@us.navy.mil}

\author[0000-0002-5187-7107]{Wendy M. Peters}
\affiliation{Naval Research Laboratory,
4555 Overlook Avenue SW,
Washington, DC 20375, USA}
\email{wendy.m.peters8.civ@us.navy.mil}

\author[0000-0003-3272-9237]{Emil Polisensky}
\affiliation{Naval Research Laboratory,
4555 Overlook Avenue SW,
Washington, DC 20375, USA}
\email{emil.j.polisensky.civ@us.navy.mil}

\author[0000-0003-3924-243X]{Jackie Villadsen}
\affiliation{Department of Physics \& Astronomy, Bucknell University, Lewisburg, PA, USA}
\email{jrv012@bucknell.edu}

\author[0000-0003-0454-3718]{Jordan M. Stone}
\affiliation{Department of Physics \& Astronomy, University of Wyoming,
1000 E. University Avenue,
Laramie, WY 82070, USA}
\affiliation{Naval Research Laboratory,
4555 Overlook Avenue SW,
Washington, DC 20375, USA}
\email{jstone30@uwyo.edu}

\begin{abstract}

Magnetically driven phenomena such as flaring events and aurorae lead ultracool dwarfs to emit at radio frequencies. Despite decades of scrutiny, a comprehensive physical understanding of their radio emission at different frequencies remains elusive, spurring on additional study of these complex objects. The VLA Low-band Ionosphere and Transient Experiment (VLITE) is a commensal instrument operating at 340 MHz on the Very Large Array. A key advantage of 340 MHz observations is their sensitivity to circumstellar disks and planets at understudied distances from the stellar disk, intermediate between GHz and low MHz sensitivities. Hard-to-find coronal mass ejections are also predicted to be detectable at 340 MHz. However, this frequency regime is relatively unprobed in ultracool dwarf studies, with few searches and no published detections to date. Here we highlight our investigation of the nearby M7-M7 binary, EI Cancri. \EIAB is magnetically active, yet has an uncharacteristically long 83-day candidate rotation period within the system. With the VLITE detection of the EI Cancri system, we present the first ever detection of ultracool dwarf emission at 340 MHz.
\end{abstract}

\section{Introduction} \label{sec:intro}

\subsection{Overview}

Ultracool dwarfs (UCDs) are comprised of brown dwarfs and the smallest main sequence stars, with spectral types M7 and later \citep{Kirkpatrick1997}. These objects are fully convective and therefore lack a tachocline, yet much of our understanding of magnetic processes in solar-type stars has hinged on this structure. Although this has left a gap in our understanding of the drivers of magnetic phenomena in UCDs, recent studies have begun to reveal the dynamos and processes behind UCD magnetic activity \citep[e.g.,][]{Williams2014,Cook2014,Kao2016}.
Overviews and recent studies of dynamos and magnetic fields in UCDs, including those in radio, can also be found in, e.g., \cite{Tang2022,Kavanagh2024,Couperus2025}. 

The vast majority of radio studies of UCDs have been at GHz frequencies \citep[e.g.,][]{Berger2006,Route2013,Kao2019,Pineda2023,Ortiz-Ceballos2024}, with additional focus at $\lesssim$200~MHz \citep[e.g.,][]{Callingham2021,Huang2024_GLEAM-X} using the LOw-Frequency ARray
\citep[LOFAR;][]{vanHaarlem2013_LOFAR} and Murchison Widefield Array \citep[GaLactic and Extragalactic All-sky Murchison Widefield Array survey eXtended; GLEAM-X;][]{Hurley-Walker2022}. Fewer studies have been performed at $\sim$340 MHz, the main frequency we examine in this study. \cite{Jaeger2011} observed two UCDs, TVLM 513-46546 and 2MASS J0036+182110, with the VLA at P-band (224-480 MHz), centered at $\sim$330~MHz. Observations spanned multiple rotations of the UCDs and resulted in non-detections.  \cite{CrosleyOsten2018a, CrosleyOsten2018b}
searched for emission from M dwarf binary system EQ Peg also in the VLA P-band and found two events in 44 hours.  \cite{VilladsenHallinan2019} used the VLA to search for coherent emission across a range of frequencies including P-band from 5 active M dwarfs, and found that emission was less likely to be detected in P-band than at GHz frequencies. The paucity of 340 MHz frequency observations to date inherently increases the value of additional studies, with the possibility of detecting orbiting bodies, coronal mass ejections, or flares in areas of phase space unprobed by those corresponding to $\lesssim$200 MHz and GHz frequencies. 

Analogously to the Jupiter system, interaction between the stellar magnetic field and an orbiting planet or plasma torus can drive radio emission \cite[see reviews by][]{Callingham2024,Murphy2025}. Observations at 340 MHz are valuable for their sensitivity to these bodies at distances from the star not covered by other frequencies \citep{Dulk1985}. Whereas GHz-frequency emission is mainly tied to the low stellar corona and its small-scale magnetic loops, 340 MHz corresponds to the further-out large-scale magnetic field at up to about 2 stellar radii \citep[see Section~\ref{ssec:eicancri_vlite} and, e.g.,][]{VilladsenHallinan2019}. This is in contrast to even lower frequencies, which probe the large-scale magnetic field at even greater radial distances \citep[$>2~R_*$ at 150 MHz;][]{Vedantham2020}. Our 340~MHz observations help fill the ``gap'' between the stellar surface and distances probed at lower frequencies; if we are blind at this frequency, we may overlook a type of radio burst specific to this frequency range, since different frequencies probe different distances from the stellar surface and different magnetic phenomena \citep{Bastian1998}. Such detections are key to holistically characterizing the system, with the potential to measure an exoplanetary magnetic field for the first time.

Probing new frequencies can also reveal the diversity of stellar radio burst types and help build towards a classification scheme that includes space weather events;  many solar radio bursts are narrow-band, with different burst types dominating at different frequency ranges \citep{Bastian1998}. \cite{VilladsenHallinan2019} describe that it is possible to detect Type II or Type III bursts at 340~MHz. These are tied to solar coronal mass ejections (CMEs) or flares in the Sun \citep{Vrsnak2008}. In other stars, we expect similar emission corresponding to eruptive events propagating through the stellar plasma.  \cite{Zic2020} and \cite{Mohan2024} also describe a mechanism for CMEs to present in low-mass stars as Type IV bursts. In other stars, we expect similar emission corresponding to eruptive events propagating through the stellar plasma; the first such stellar Type II candidate was recently observed by \cite{Callingham2025} at 120–165 MHz. \cite{Zic2020} and \cite{Mohan2024} also present sub-GHz observations of stellar Type IV bursts, which in the Sun are well-correlated with post-eruptive coronal reconfiguration. These results motivate continued searches at low and intermediate MHz frequencies to detect additional scarce signatures of CMEs. 

In this paper, we report on the discovery of 340 MHz emission from UCD binary \EI AB, marking the first detection of an UCD at this frequency. We demonstrate multi-frequency radio emission in this system spanning over three decades. First we introduce the \EI system in Section \ref{ssec:EICancri}. In Section \ref{sec:observations}, we describe the observations of \EI using VLITE and results. In Section~\ref{sec:discussion}, we discuss the nature of the detected emission. Lastly, in Section~\ref{sec:conclusion}, we draw our conclusions and describe prospects for future work on this system.

\subsection{The EI Cancri System} \label{ssec:EICancri}

\EIAB (G~9-38AB, GJ~1116~AB, TIC~197251248 (A) + TIC~471012520 (B)) was first identified as a flare star system by \cite{Pettersen1985}. The system consists of two main sequence M7 stars \citep{Newton2014} at $\sim$5.12 pc \citep{Bailer-Jones2021}, with masses of 0.120$\pm$0.014~M$_\odot$ and 0.103$\pm$0.014~M$_\odot$ \citep{Winters2021} and at a separation of $2.5478\arcsec\pm0.0003\arcsec$ \citep{Tokovinin2020}. Multiple measurements of rotation period have been taken for the system (83 days by \cite{Newton2016} and $\sim10$ hours by \cite{Jeffers2018}). However, because the measurements were obtained from unresolved observations, neither period can be conclusively attributed to a specific component (see Section~\ref{ssec:mag_rot}).

The orbital parameters of \EI have not yet been determined. The system's projected angular separation of $2.5478\arcsec\pm0.0003\arcsec$ \citep{Tokovinin2020} corresponds to a plane-of-sky separation of 13~AU. In contrast, the individual stellar parallax measurements from Gaia~DR3 suggest a line-of-sight separation as large as $\sim$10,000 AU \citep{GaiaCollaboration2023_DR3, Bailer-Jones2021}. However, this larger value is highly uncertain, as parallax-derived distances for close, bright binary stars can be unreliable due to poor astrometric fits \citep{El-Badry2021}. Even assuming the conservative 13~AU minimum physical separation, strong magnetospheric interaction between the components is unlikely, which is consistent with findings for other radio-emitting UCD binaries \citep{Kao2025}. The fundamental properties of the \EI system are summarized in Table 1.

\begin{deluxetable}{c|cc|cc|c}[!h]
\label{table:BasicInfo}
\tabletypesize{\footnotesize}
\tablecaption{Basic Information}
\setlength{\tabcolsep}{0.03in}
\tablewidth{0pt}
\tablehead{
\colhead{Property} &
\colhead{EI Cancri A} &
\colhead{Error} &
\colhead{EI Cancri B} &
\colhead{Error} &
\colhead{Reference}
}
\startdata
\hline
\hline
Common Names & GJ~1116 A & \nodata & GJ~1116~B & \nodata & \nodata \\
 & G~9-38A & \nodata & G~9-38B & \nodata & \nodata \\ 
 & TIC~197251248 & \nodata & TIC~471012520 & \nodata & \nodata \\ 
\hline
\multicolumn{6}{c}{Astrometry and Kinematics} \\
\hline
RA (epoch 2016.0; hms) & $08^d58^m14.206^s$  & $<0.001^s$ & $08^d58^m14.086^s$ & $<0.001^s$ & Gaia DR3 \\ 
Dec (epoch 2016.0; dms) & $+19^\circ45\arcmin46.655\arcsec$ & $<0.001\arcsec$ & $+19^\circ45\arcmin45.284\arcsec$ & $<0.001\arcsec$ & Gaia DR3 \\ 
Parallax (mas) & 194.1443 & 0.1228 & 196.2619 & 0.1976 & Gaia DR3 \\
Distance (pc) & 5.1499 & $^{+0.0027}_{-0.0039}$ & 5.0939 & $^{+0.0062}_{-0.0051}$ &\cite{Bailer-Jones2021} \\
Total Proper Motion (mas yr$^{-1}$) & 773.574 & \nodata & 937.770 & \nodata & Gaia DR3 \\
RA Proper Motion (mas yr$^{-1}$) & -767.060 & 0.122 & -937.133 & 0.190 & Gaia DR3 \\
Dec Proper Motion (mas yr$^{-1}$) & -100.176 & 0.083 & -34.559 & 0.138 & Gaia DR3 \\
Gaia RUWE & 2.574 & \nodata & 3.539 & \nodata & Gaia DR3 \\
AB Separation (2019.9474; arcsec) & 2.5478 & 0.0003 & \nodata & \nodata & \cite{Tokovinin2020} \\
AB Position Angle (2019.9474; deg) & 244.8 & 0.3 mas & \nodata & \nodata & \cite{Tokovinin2020} \\
\hline
\multicolumn{6}{c}{Photometry} \\
\hline
Gaia G (mag) & 11.9663 & 0.0029 & 12.4856 & 0.0030 & Gaia DR3 \\
Gaia BP (mag) & 14.3240 & 0.0045 & 15.0630 & 0.0054 & Gaia DR3 \\
Gaia RP (mag) & 10.5470 & 0.0058 & 11.0246 & 0.0068 & Gaia DR3 \\
2MASS K$_s$ (mag) & 6.889 & 0.023 & \nodata & \nodata & 2MASS; combined \\
GALEX NUV (mag) & 19.4572 & 0.1112 & \nodata & \nodata & \cite{Bianchi2017_GALEX}; combined \\
GALEX FUV (mag) & 20.7616 & 0.2886 & \nodata & \nodata & \cite{Bianchi2017_GALEX}; combined \\
ROSAT flux ($10^{-12}$ erg cm$^{-2}$ s$^{-1}$) & 2.9770 & \nodata & \nodata & \nodata & \cite{Boller2016_ROSAT_2RXS}; combined \\ 
L$_{X}$ (ROSAT; $10^{27}$ erg s$^{-1}$) & 9.2647 & \nodata & \nodata & \nodata & \cite{Boller2016_ROSAT_2RXS}; combined \\ 
\hline
\multicolumn{6}{c}{Derived Properties} \\
\hline
Spectral Type & M7 & \nodata & M7 & \nodata & \cite{Newton2014} \\
Mass (M$_\odot$) & 0.120 & 0.014 & 0.103 & 0.014 & \cite{Winters2021} \\
Radius (R$_\odot$) & 0.148 & 0.013 & 0.132 & 0.013 & \cite{Pineda2021} M-R Relation$^\dagger$ \\
EW$_{H\alpha}$ (\AA)  & -5.719 & 0.044 & \nodata & \nodata & \cite{Newton2017}; combined \\
L$_{H\alpha}$/L$_{bol}$ & $ 1.027 \times 10^{-4}$ & \nodata & \nodata & \nodata & \cite{Newton2017}; combined \\
P$_{rot,combined}$ (days)$^\ddagger$ & 83.270 & \nodata & \nodata & \nodata & \cite{Newton2016}; combined \\
P$_{rot}$ (days)$^\ddagger$ & 113 & \nodata & 116 & \nodata & \cite{Newton2017} eqn. 6 \\
vsini (km~s$^{-1}$)$^\ddagger$ & 16.70 & 0.64 & \nodata & \nodata & \cite{Jeffers2018}; combined \\
\hline
\hline
\enddata
\tablenotetext{\dagger}{\url{https://github.com/jspineda/stellarprop}}
\tablenotetext{\ddagger}{Note a discrepancy between periods from \cite{Newton2016,Newton2017} and from \cite{Jeffers2018}; see Section~\ref{ssec:mag_rot}.}
\end{deluxetable}

While no exoplanet candidates have been reported to date, the Gaia Data Release 3 \citep[DR3;][]{GaiaCollaboration2016,GaiaCollaboration2023_DR3,Babusiaux2023_GaiaDR3_Validation} Renormalised Unit Weight Error (RUWE) values for both components exceed $\gtrsim1.4$, potentially indicating unresolved astrometric perturbations consistent with an additional companion \citep{Lindegren2018,Belokurov2020,Stassun2021,10.5270/esa-qa4lep3}. However, it is possible that these higher RUWE values stem from the exclusion of an orbital fit in the astrometric solution for these two stars, rather than a third body. Some orbital motion has been seen in Southern Astrophysical Research (SOAR) Telescope  speckle data over 2019-2021 \citep[][Eliot Vrijmoet, private communication]{Tokovinin2020, Vrijmoet2022}. The system increased by 0.1$\arcsec$ in separation and rotated 3.6$^\circ$ in position angle; this may be enough motion to account for the high RUWE in both components in Gaia DR3. Speckle observations taken by the aforementioned SOAR speckle team and Robo-AO team \citep{Salama2022} have yielded no discovery of a third member of the system to date. 

The Transiting Exoplanet Survey Satellite \cite[TESS;][]{Ricker2015_TESS} lightcurve of the \EI system shows spot modulation and frequent flares. With a TESS resolution of 21$\arcsec$ and \EIAB a near-equal-mass binary, flux cannot be disentangled between the two stars, so the source(s) of these magnetic activity signatures cannot be distinguished between them. As we'll discuss in Section~\ref{sec:discussion}, magnetic phenomena can lead to other types of emission, such as excess UV and X-ray flux (see Table~\ref{table:BasicInfo}), in addition to the radio emission presented in this work.

Previous detections of radio emission from \EIAB have been made from 855.5 MHz to 5 GHz and 90-220 GHz, as shown in Table~\ref{table:ArchivalRadio}. However, no known studies have been performed to interpret these data and put them in context with the known magnetic properties of the system.

Here we deepen the radio characterization of this system by extending down to the yet unexplored 340~MHz regime, revealing the first UCD detections at this frequency and interpreting the cumulative radio detections of this system to date.

\section{Observations and Results} \label{sec:observations}

\subsection{VLITE}

The VLA Low-band Ionosphere and Transient Experiment (VLITE)\footnote{\url{https://vlite.nrao.edu}} is a commensal instrument on the Karl G. Jansky Very Large Array (VLA). Operating at a central frequency of 340 MHz, with an effective 40 MHz bandwidth, it records and correlates data during nearly all standard VLA observations at GHz frequencies, also called the primary frequencies. Science operations began with 10 antennas in 2014 November, and the system currently operates on 18 antennas. VLITE collects more than 6000 hrs/yr of data. The wide field of view ($2
.45^{o}$ FWHM) provides measurements of sources over a large sky area around each primary science target position. 

Data are typically stored in blocks containing all data acquired within a 24-hour period. Due to its commensal nature, the VLITE database has a non-uniform distribution of depth of sky coverage. Data are processed on a daily basis, with each primary observing frequency processed separately due to variations in the instrumental response. A dedicated calibration and imaging pipeline relies on a series of standard tasks in the \texttt{Obit} \citep{Cotton2008_Obit} and \texttt{Astronomical Image Processing System (AIPS)} \citep{Greisen2003_AIPS} software packages. Correction of delays, frequency-dependent complex gains, and frequency-independent complex gains are made using observations of one of six primary calibrators (3C48, 3C138, 3C147, 3C286, 3C295, and or 3C380), using models which are set to the \cite{Perley2017} flux scale. After editing to remove radio frequency interference (RFI), all data are phase calibrated to a sky model based on the National Radio Astronomy Observatory (NRAO) VLA Sky Survey \citep[NVSS;][]{Condon1998_NVSS}. The data are imaged using the \texttt{Obit} task `MFImage.'  Calibrated visibilities and final images are stored in archives at the Naval Research Laboratory (NRL). For more details on standard VLITE processing, please see \citet{Polisensky2016}.

Source detection is performed using the \texttt{VLITE Database Pipeline} \citep[VDP;][]{vdp}, which utilizes \texttt{PyBDSF} \citep{Mohan2015} source extraction to catalog sources with signal-to-noise ratios (S/N) $>5$. VDP calculates and applies calibration for VLITE's elaborate primary beam response \citep{poli2024} and stores end data products in a Structured Query Language (SQL) database.

Key to the current study is deployment of a ``time-chopping'' methodology to sample the VLITE data at different durations and cadences. 
Using the same software tools used in the imaging pipeline, we ``chop'' the data into a series of shorter-time images from the larger whole. This chopping of long-duration images (the longest being $\sim7$ hours) into shorter cadences enables us to search for shorter events.

\subsection{EI Cancri in VLITE}
\label{ssec:eicancri_vlite}

We searched the VLITE archive for emission from known radio stars in the Sydney Radio Star Catalogue \citep{Driessen2024_SRSC}. From the detected stars, we chose the UCD binary system \EI for additional study because the detection dataset is at VLITE’s highest resolution and includes sufficient data to search for bursts of emission on timescales of minutes and hours over the course of a day.

In addition, \EI is the only UCD system detected using VLITE's routine pipeline amongst the ~3600 stars and brown dwarfs known within 20 pc \citep{Kirkpatrick2024}.
To search for radio emission resolved between the binary components of EI Cancri, we focus on A-configuration observations, which provide the highest angular resolution available ($\sim5^{''}$). Since the 2017 upgrade to 18 antennas that enabled VLITE imaging in A configuration, 111 images include EI Cancri within the field of view. These observations span 2018 March 2 to 2024 December 19. During this time, the \EI system moves 8.7$\arcsec$ across the sky; we use Gaia DR3 proper-motion-shifted coordinates for the \EI system at the midpoint time of the VLITE observations (RA = $08^h58^m13.81^s$, Dec = $+19^\circ45\arcmin45.59\arcsec$, epoch = 2021.62839). All observations of \EI are 0.874$^\circ$ from pointing center, which corresponds to a common calibrator, J0854+2006. There are no data in which \EI is the intended target of the Cassegrain antennae observers; these are serendipitous detections outside the field of view of the primary GHz frequency observations.

\begin{deluxetable}{lccccccccc}[!h]
\label{table:ArchivalRadio}
\tabletypesize{\scriptsize}
\tablecaption{Archival Radio Data}
\setlength{\tabcolsep}{0.03in}
\tablewidth{0pt}
\tablehead{
Survey/Facility &
Freq. &
Resolution & 
Component &
Flux Density &
Error &
Epoch &
Ref &
Brightness Temp. &
Error
\\
\nodata & (MHz) & \nodata & \nodata & 
(mJy) & (mJy) & 
(yr) & \nodata & 
(This work; K) &
(This work; K)
}
\startdata
\hline
\hline
NVSS$^*$   & 1400                      & 45\arcsec & U     & 3.3     & 0.4      & 1993.833     & C98      & \nodata & \nodata \\ 
FIRST$^\dagger$  & 1400                      & 5\arcsec & U      & 3.67    & 0.151    & 1998.775         & B95, H15 & \nodata & \nodata\\ 
VLA  & 5000                      & $\sim$1\arcsec-10\arcsec & U      & 1.482   & 0.082    & 2005.671233         & B09 & \nodata & \nodata\\ 
\hline 
VLITE & 340 & $\sim$4$\arcsec$ & U & 2.70 & 0.35 & 2018.3152329 & this work & \nodata & \nodata \\
\hline 
ACT & 220 GHz          & 1\arcmin & U     & 103     & 31       & 2018.8718379 & L23      & \nodata &  \nodata \\ 
ACT & 150 GHz          & 1\arcmin & U     & 197     & 9        & 2018.8718379 & L23      & \nodata &  \nodata\\ 
ACT & 90 GHz           & 1\arcmin & U     & 269     & 9        & 2018.8718379 & L23      & \nodata & \nodata \\ 
\hline 
VLASS 1 & 3000              & $\sim2.5\arcsec$ & A    & 0.82 & 0.18 & 2019.2822444     & L20, this work      & 5.50$\times 10^{11}$ & 1.55$\times 10^{11}$\\ 
VLASS 1 & 3000              & $\sim2.5\arcsec$ & B    & 1.082 & 0.189 & 2019.2822444     & L20      & 8.92$\times 10^{11}$ & 2.35$\times 10^{11}$\\ 
\hline
ASKAP $\|$ 02-03hour & 855.5      & $\sim15\arcsec$ & U     & 6.87    & 0.05     & 2020.2936        & M20, Dr24      & \nodata & \nodata \\
ASKAP & 855.5                    & $\sim15\arcsec$ & U     & 1.45    & 0.02     & 2020.2938        & M20, Dr24      & \nodata & \nodata \\ 
ASKAP  $\|$ RACS-MID & 1367.5     & $\sim10\arcsec$ & U     & 2.2     & 0.5      & 2021.0293        & M20, Du24, Dr24      & \nodata & \nodata \\ 
ASKAP & 1367.5                   & $\sim10\arcsec$ & U     & 3.7     & 0.5      & 2021.0431        & M20, Dr24      & \nodata & \nodata \\ 
\hline
VLASS 2 & 3000            & $\sim2.5\arcsec$ & A & 1.269 & 0.130  & 2021.877854    & L20      & 8.51$\times 10^{11}$ & 1.73$\times 10^{11}$\\
VLASS 2 & 3000            & $\sim2.5\arcsec$ & B & 1.092 & 0.126 & 2021.877854    & L20      & 9.01$\times 10^{11}$ & 2.06$\times 10^{11}$\\ 
\hline
ASKAP $\|$ RACS-HIGH & 1655.5      & $\sim8\arcsec$  & U     & 1.4     & 0.3      & 2021.9940        & M20, Dr24, Du25      & \nodata & \nodata \\ 
ASKAP & 855.5                    & $\sim15\arcsec$ & U     & 2.6     & 0.2      & 2022.4066        & M20, Dr24      & \nodata & \nodata \\ 
ASKAP & 855.5                    & $\sim15\arcsec$ & U     & 1.5     & 0.2      & 2022.4093        & M20, Dr24      & \nodata & \nodata \\ 
ASKAP & 943.5                    & $\sim15\arcsec$ & U     & 5.2     & 0.4      & 2023.9886        & M20, Dr24      & \nodata & \nodata \\ 
ASKAP & 943.5                    & $\sim15\arcsec$ & U     & 5.6     & 0.4      & 2023.9912        & M20, Dr24      & \nodata & \nodata \\ 
ASKAP & 943.5                    & $\sim15\arcsec$ & U     & 4.1     & 0.1      & 2024.0265        & M20, Dr24      & \nodata & \nodata \\ 
ASKAP & 1367.5                   & $\sim10\arcsec$ & U     & 2.42    & 0.07     & 2024.1602        & M20, Dr24      & \nodata & \nodata \\ 
\hline
VLASS 3 & 3000            & $\sim2.5\arcsec$ & A &  1.413 & 0.161  & 2024.333561 & L20, this work      & 9.48 $\times 10^{11}$ & 1.98$\times 10^{11}$ \\ 
VLASS 3 & 3000            & $\sim2.5\arcsec$ & B &  1.706 & 0.165  & 2024.333561 & L20, this work      & 1.41 $\times 10^{12}$ & 3.09$\times 10^{11}$\\ 
\hline
\hline
\enddata
\tablenotetext{*}{Note that there is a $\sim7\arcsec$ separation between NVSS and Gaia proper-motion shifted coordinates. This separation is mostly in Dec and comparable to the NVSS Dec error of 5.5\arcsec. C98 highlight a lower reliability in optically identifying sources weaker than 5 mJy and demonstrate that position uncertainties smaller than 30$\arcsec$ have a false alarm probability of $\sim0.01$. We believe this detection likely corresponds to emission in the \EI system, but caution the reader of these uncertainties.}
\tablenotetext{}{U = unclear whether the flux comes from \EI A, B or both.}
\tablenotetext{}{NVSS: National Radio Astronomy Observatory (NRAO) Very Large Array (VLA) Sky Survey, FIRST: Faint Images of the Radio Sky at Twenty-centimeters, ACT: Atacama Cosmology Telescope, VLASS: VLA Sky Survey, ASKAP: Australian Square Kilometre Array Pathfinder, RACS: Rapid ASKAP Continuum Survey}
\tablenotetext{}{C98=\cite{Condon1998_NVSS},
B95=\cite{Becker1995_FIRST},
H15=\cite{Helfand2015_FIRSTcatalog},
B09=\cite{Bower2009},
L23=\cite{Li2023_ACT},
L20=\cite{Lacy2020_VLASS},
M20=\cite{McConnell2020},
Dr24=\cite{Driessen2024_SRSC} \& \url{https://radiostars.org/},
Du24=\cite{Duchesne2024},
Du25=\cite{Duchesne2025}}
\tablenotetext{\dagger}{Strangely, FIRST emission is detected between the proper-motion-corrected coordinates of each star, but is point-like rather than elongated, suggesting that it comes from only one source.}
\end{deluxetable}

\begin{figure}
\begin{center}
    \includegraphics[width=\textwidth]{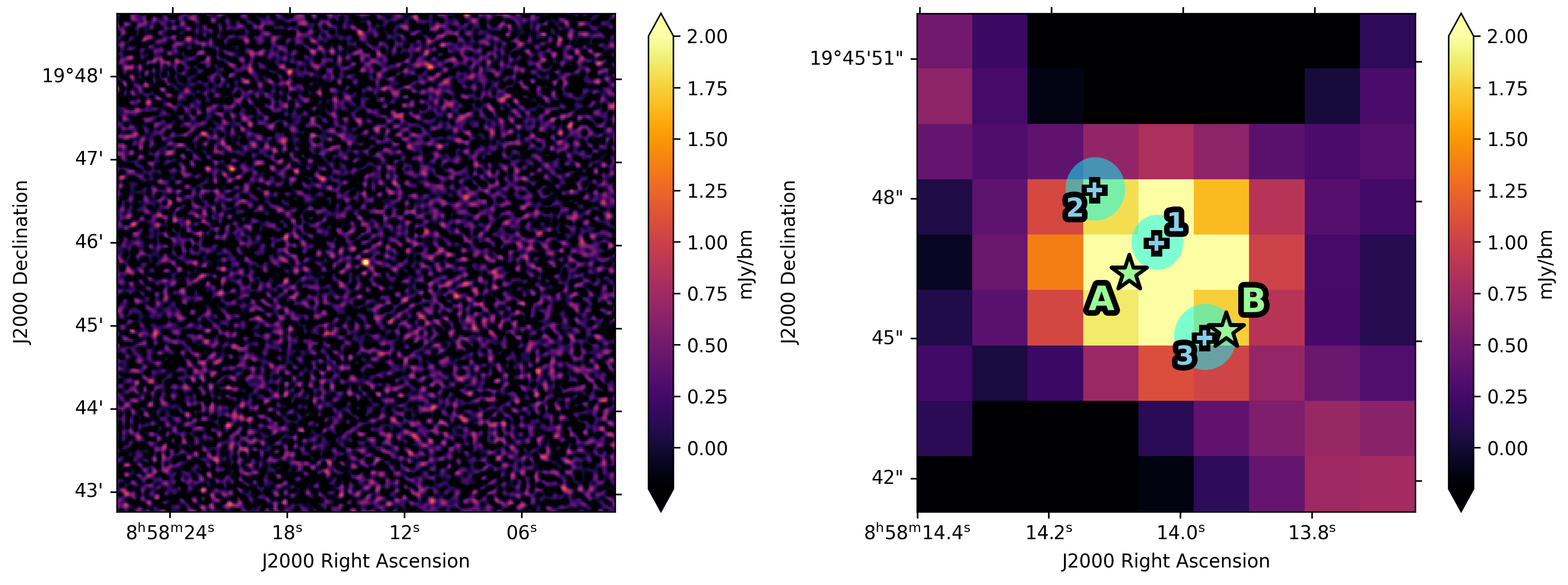} 
    \caption{Left: VLITE detection of EI Cancri on 2018 April 26 from an image with 7 hours on source across a 28-hour duration. The image has a pixel scale of 1.2$\arcsec$ pixel$^{-1}$ and a resolution of $3.9^{''} \times 3.8^{''}$. The stellar binary is prominently detected with a peak flux of 2.7$\pm$0.4 mJy bm$^{-1}$ and S/N = 7.71. Right: Zoom-in on the detection. The positions of the stellar components are shown by the star symbols, corrected for proper motion. The positions of the three individual detections in the time-chopped (10 minute) images are marked with `+' symbols, with shaded light blue circles representing 1-$\sigma$ positional errors. Table~\ref{table:ArchivalRadio} places the 7-hour event in context with other archival detections, and the three 10-minute events are further described in Table~\ref{table:VLITEDetections}.}
\label{fig:VLITEdetection}
\end{center}
\end{figure}

Emission from \EI at 340 MHz is detected in a deep archival image from VLITE observations taken during project GG083, which observed the nearby blazar OJ 287 at 22 GHz. The VLITE image of EI~Cancri, shown in the left panel of Figure~\ref{fig:VLITEdetection}, includes 7.036 hours of on-source time, spread over a total of 28 hours on 2018 April 26 to 2018 April 27 (see also Table~\ref{table:ArchivalRadio}). The image has a resolution of $3.9^{''} \times 3.8^{''}$ at a position angle of $66^{o}$, and a local rms noise of 0.35 mJy bm$^{-1}$. To determine the astrometric accuracy of the image, we compare the measured positions of 82 compact point sources with S/N $>$ 10 to their positions in the 1400 MHz FIRST survey \citep{Becker1995_FIRST}, which has a comparable resolution ($5^{''}$) to the VLITE image and a measured astrometric accuracy of $< 0.05^{''}$.  The median astrometric accuracy of the VLITE sources relative to FIRST is $0.04^{''}$, comparable to the survey accuracy.

The right panel of Figure~\ref{fig:VLITEdetection} shows a small region of the image centered on the detected radio emission. The positions of \EI A and B at the time of the observation are labeled. The VLITE source centroid is closest to \EI A, with an offset of 0.64$^{''}$ while the offset to \EI B is 2.0$5^{''}$. The shape of the time-averaged position measured in this long observation matches that of a point source rather than an unresolved binary, as would be expected if the emission was uniformly distributed between \EI A and B. We do not have strong reason to believe there is a third UCD companion in the system generating the emission (see Section~\ref{ssec:EICancri}). Given the resolution of $\sim$ 4$^{''}$ and our measured astrometric accuracy, the VLITE position is most consistent with an association with \EI A. 

Due to the prevalence of shorter-timescale optical flares, the long 7-hour image likely contains multiple shorter outbursts; therefore, to search for time-resolved radio emission and identify shorter events, we split the full 111-image dataset into smaller time bins. With no UCD detections in this frequency to date, we guide our cadence selection using optical flare timescales, even if there is some uncertainty in their physical connection (i.e. radio emission may not be from flares). We choose a 10-minute cadence, in line with findings by \cite{Pettersen1985} of about five flares per hour. The result is a total of 574 images, totaling $\sim$23 hr on source. Each image was processed using VDP in a non-standard mode that catalogs sources with S/N $>3$ to identify potential low-significance detections of EI~Cancri. 

We identified 42 sources within $10^{\prime\prime}$ of the proper motion-corrected position of EI~Cancri. At these low S/N levels, spurious noise detections are a concern. To quantify the likelihood of false associations, we conducted a statistical analysis assuming all detections are randomly distributed artifacts. For each image, the expected number of false positives is the product of the $10^{\prime\prime}$ association solid angle $\Omega_*$ and the areal density of detections, $n_i / \Omega_{\mathrm{FoV}}$, where $n_i$ is the number of sources and $\Omega_{\mathrm{FoV}} = 0.25$ deg$^2$ is the image field of view. Summing over all $N$ images yields the expected number of false associations:

\begin{equation}
\lambda = \sum_{i=1}^{N} \Omega_*  \frac{n_i}{\Omega_{\mathrm{FoV}}}
\end{equation}

We detected 38 associations with S/N between 3 and 4, consistent with the expected 38.7 false positives. For S/N~$>4$, we detected 4 associations, compared to 2.1 expected false positives. Visual inspection of the four high-S/N candidates revealed one likely artifact. The remaining three significant events all occurred on 2018 April 27, during the 7-hour interval when \EI was detected. 

To confirm the events, search for the times of highest S/N, and map out the light curve, we applied boxcar smoothing to the 7-hour dataset, such that each 10-minute image overlaps the next by 9 minutes. Each bin included at least 6.5 minutes of time on-source, with the majority containing a full 10 minutes. With this methodology, we are able to identify the time with the highest S/N for each event, presented in Table~\ref{table:VLITEDetections}, with positions displayed in the right panel of Figure~\ref{fig:VLITEdetection}. These three events have peak emissions of 14.4 mJy, 10.1 mJy, and 10.3 mJy at UT = 00:09, 02:48, and 03:41, respectively. In all cases, the emission is spatially unresolved. We adopt the fitted peak brightness as the source flux density, because the fitted sizes and total fluxes of weak sources in VLITE data are biased to larger values \citep{poli2024}.

\begin{deluxetable}{cccccccccccc}[!h]
\label{table:VLITEDetections}
\tablecaption{VLITE Detections}
\setlength{\tabcolsep}{0.03in}
\tablewidth{0pt}
\tablehead{
Event \# &
Date and Midtime &
RA &
$\sigma_{RA}$ &
Dec &
$\sigma_{Dec}$ &
Flux Density  &
Error &
Time On-source &
S/N &
$T_{B}$ &
Error
\\
\nodata & \nodata & (hms) & (arcsec) & (dms) & (arcsec) & (mJy) &  (mJy) & 
(sec) & \nodata & 
($10^{12}$ K) & ($10^{12}$ K)
}
\startdata
\hline
\hline
1 & 2018-04-27 UT 00:09:00  & $08^h58^m14.04^s$ & 0.25 & $+19^\circ45^\prime47.11^{\prime\prime}$ & 0.20 &  14.4  &  3.2  &  600  &  5.7  &  9.7  &  2.7 \\
2 & 2018-04-27 UT 02:48:00  & $08^h58^m14.14^s$ & 0.34 & $+19^\circ45^\prime48.12^{\prime\prime}$ & 0.42 &  10.1  &  2.5  &  600  &  4.9  &  6.8  &  2.1 \\
3 & 2018-04-27 UT 03:41:00  & $08^h58^m13.96^s$ & 0.27 & $+19^\circ45^\prime44.73^{\prime\prime}$ & 0.47 &  10.3  &  2.6  &  600  &  4.9  &  6.9  &  2.1 \\
\hline
\hline
\enddata
\end{deluxetable}

At low radio frequencies, the ionosphere introduces a refractive phase error which can be removed by calibrating to a sky model with known source positions; for VLITE we use the NVSS for this purpose \citep{Polisensky2016}.  However, on baselines longer than the projected scale of structures in the ionosphere, the field-of-view over which this position-independent calibration is valid, or isoplanatic patch,  may be less than the desired area of the image. In VLITE archival data we see small position-dependent offsets only in the largest VLA configuration (A configuration). Generally the isoplanatic patch is $\gtrsim 1.0^{o}$, and typical errors are $\lesssim2-3^{''}$ on timescales of minutes. Over longer observations, such as the 7 hours of data in the original archival detection image, the net result of leaving the direction-dependent ionospheric phases uncorrected is to slightly smear the sources, leading to a decrease in the peak flux, but an accurate total flux and position \citep{cohen2003}. 

Unfortunately, 10 minutes is not sufficient to average out the ionospheric effects, so the astrometric uncertainty is expected to be larger than in the 7 hour image analyzed previously. Due to the decreased sensitivity of the shorter images, we do not detect enough sources to robustly analyze the astrometric errors in each image individually as we did for the 7 hour image. Instead, we estimate the position errors by measuring the position shifts of 12 point sources within a radius of $0.5^{o}$ of \EI in all of the 10 minute images, for a total of 517 detections. The $1\sigma$ uncertainty is $0.53^{''}$ in RA and $0.38^{''}$ in Dec, with a small number of outliers, roughly $1\%$ of the total, at offsets $\sim 2^{''}$. 

In the right panel of Figure~\ref{fig:VLITEdetection}, the positions of the three ten-minute detections are indicated with blue crosses. A shaded region is shown around each detection, representing the 1$\sigma$ positional uncertainty derived from the fitting uncertainty, ionospheric positional variations, and astrometric calibration errors, added in quadrature. Each detection has a point-like profile. The position of burst 3 matches that of \EI B within the estimated errors. Bursts 1 and 2 are both closer to \EI A, although neither matches the star's position within the estimated uncertainty. If both stars are emitting then it would explain the relatively central position of the 7-hour detection between \EI A and B. Additional follow-up observations at this frequency with direction-dependent calibration would be necessary to confirm whether both stars are active as suggested by these results. The source coordinates and their 1$\sigma$ positional uncertainties, derived from the Gaussian fits, are listed in Table~\ref{table:VLITEDetections} alongside the times and flux densities.

To determine the nature of the emission, we calculate the stellar disk-averaged brightness temperature ($T_B$) of each event (henceforth ``brightness temperature''). Brightness temperature in K is determined as a function of flux density ($S$), distance to the star ($d$), and size of the emitting region ($r$): 

\begin{equation}
T_B = 2.8156 \times 10^{-7} \cdot S \left(\frac{d}{r}\right)^2
\end{equation}
for $S$ in mJy, $\frac{d}{r}$ unitless, and adoption of the stellar radius of \EI A in units of $R_\odot$ for $r$. We estimate stellar radius by substituting masses from \cite{Winters2021} into the relation from \cite{Pineda2021}. Derived brightness temperatures from VLITE data $\gtrsim 10^{12}$~K indicate a coherent emission mechanism \citep{Dulk1985,Melrose1991}. All of our VLITE detections meet this criterion. However, at 340 MHz, the radius of emission could be from large magnetic loops extending far above the stellar surface. 
It is likely that the emission is coherent from the electron cyclotron maser instability (ECMI), in line with findings for other UCDs \citep{Hallinan2008,VilladsenHallinan2019}, but because brightness temperature scales smaller with a larger radius, possibly below the coherent/incoherent boundary, it is important to consider this possibility and its implications.

If the emission occurs at the fundamental cyclotron frequency, as is common for ECMI, then the observed frequency of 340 MHz corresponds to a magnetic field strength in the emission region of 120~G. If the stars' rotation periods are $\sim$80~days, then 120~G is a typical average surface magnetic field \citep{Reiners2022}. Conversely, if the stars' rotation periods are $\sim$10~hr, then their average surface fields are likely multiple kG, with dipolar fields of order 200-1000~G, as found for the rapidly rotating late-M binary UV Ceti \citep{Kochukhov2017}. If we assume a kG dipole field, this magnetic field would decay to 120~G at a radius of about 2~$R_*$. Thus, for the favored mechanism of ECMI, the emission is predicted to originate at around 1-2~$R_*$, depending on the stars' rotation and magnetic properties. If the source size has a radius of 2~$R_*$, our brightness temperatures would be a factor of 4 smaller, overlapping the $10^{12}$~K boundary differentiating coherent and incoherent emission.

If the emission is gyrosynchrotron, it can originate from weaker field strengths due to relativistic boosting of the emission frequency \citep[$\times$10-100,][]{Dulk1985}, so the emission could come from large-scale loops in the star's closed magnetic field.  The question of how large such loops can be is not well-resolved in the literature for late M dwarfs, and this is compounded by our targets' unknown rotation and magnetic properties. \cite{Vidotto2014} modeled the winds of early to mid M dwarfs based on spectropolarimetric maps and found that the closed field typically ended around 5~$R_*$. \cite{Kavanagh2021} modeled the magnetosphere of late M dwarf Proxima Centauri; their model appears to have a closed field extending to 5-10 $R_*$, although they found for early M dwarf AU Mic that the closed field radius depends to order of magnitude level on the stellar wind mass loss rate.  In contrast, UCDs with weak winds may have closed magnetic fields extending up to hundreds of stellar radii \citep{Turnpenney2017}, with one radiation belt observed extending at least $\sim$10 stellar radii from the UCD \citep{Kao2023}. Emission from \EI A or B at $\sim$10~$R_*$ would correspond to a brightness temperature 1-2 orders of magnitude below the coherent/incoherent emission boundary. We cannot definitively rule out the possibility of gyrosynchrotron emission from a large extended magnetosphere, but favor cyclotron maser due to its luminosity and the prevalence of coherent emission at lower frequencies on other M dwarfs \citep{Callingham2021}. 

Deeper discussion of possible emission mechanisms can be found in Section~\ref{ssec:mechanisms}. Table~\ref{table:VLITEDetections} reports parameters, including brightness temperature, for our three VLITE detections of \EI AB.

\subsubsection{Additional Emission at 340 MHz}\label{sssec:additional_emission}
The three VLITE detections were identified within a 7-hour image upon time chopping at a 10 minute cadence. However, we also identify a signal of S/N = 7.7 within the full, non-chopped, 7-hour image, with a peak flux of 2.7 $\pm$ 0.4 mJy corresponding to $T_B = 1.8 \times10^{12}~\text{K} \pm 4.0 \times10^{11}$ K. This begs the question of whether there is additional emission below our time-chop detection limits. Integrating the flux of the three detections across their 10-minute timescales and dividing by the full image time as in equation \ref{eq:quiescent}, we calculate an expected flux of 0.8 mJy during the 7-hour period, short of the flux from the actual 7-hour image. This result suggests additional flux below our sensitivity at the 10-minute cadence, but measurable in the combined image. Thus, we find that there is a low level of emission from \EI A and/or B during the 2018 April epoch below our chopping detection limits. 

\begin{equation}
    \label{eq:quiescent}
    f_{peak,3~detections} = \
    \frac{(f_{peak,1} + \
    f_{peak,2} + \
    f_{peak,3})~\tau_{chop}}{\tau_{~7 hr}}
    = 0.8 \text{~mJy} ~<~ f_{peak,7 hr} = 2.7 \text{~mJy}
\end{equation}

To test for this lower-level emission between the bursts, we removed the three burst time intervals from the 7-hour dataset and reimaged the remainder. We measure a source peak of $2.2 \pm 0.4$ mJy. Because of the relatively large uncertainty, this is consistent with no change from the full 7-hour source peak of $2.7 \pm 0.4$ mJy; it also agrees, within the errors, with the predicted flux decrease from removing the bursts. However, it is not consistent with all of the source flux being contained in the 3 bursts, and supports the idea that there must be some additional emission present, whether it is a series of bursts below our 10-minute detection threshold, a constant continuum emission, or a combination of both.

To search for quiescent emission beyond our 7-hour dataset, we combined images spanning 3-month intervals of VLITE data throughout the full dataset, excluding the 7-hour detection image. The time frame was chosen to be as long as possible without significant shifting due to proper motion. No emission outside the 2018 dates is detected, with $3\sigma$ limits that are typically $4 - 6$ mJy bm$^{-1}$ or higher. Thus, these combinations would not detect the 2.7 mJy emission measured in the 7-hour image from 2018. Combining the 7-hour detection image with other images taken in early 2018 lowers the measured flux, which suggests that the true quiescent level, if any, is even lower. \cite{Kao2025} find that UCD binaries are more likely to exhibit quiescent emission than single UCDs. Quiescent emission below our detection limit is unlikely to be driven by flares in the system, although the flares may populate the necessary charged particles \citep{Kao2023}. 

\subsection{Higher Frequency Emission}
\label{ssec:higherfreq}

\begin{figure}
\begin{center}
    \includegraphics[width=\textwidth]{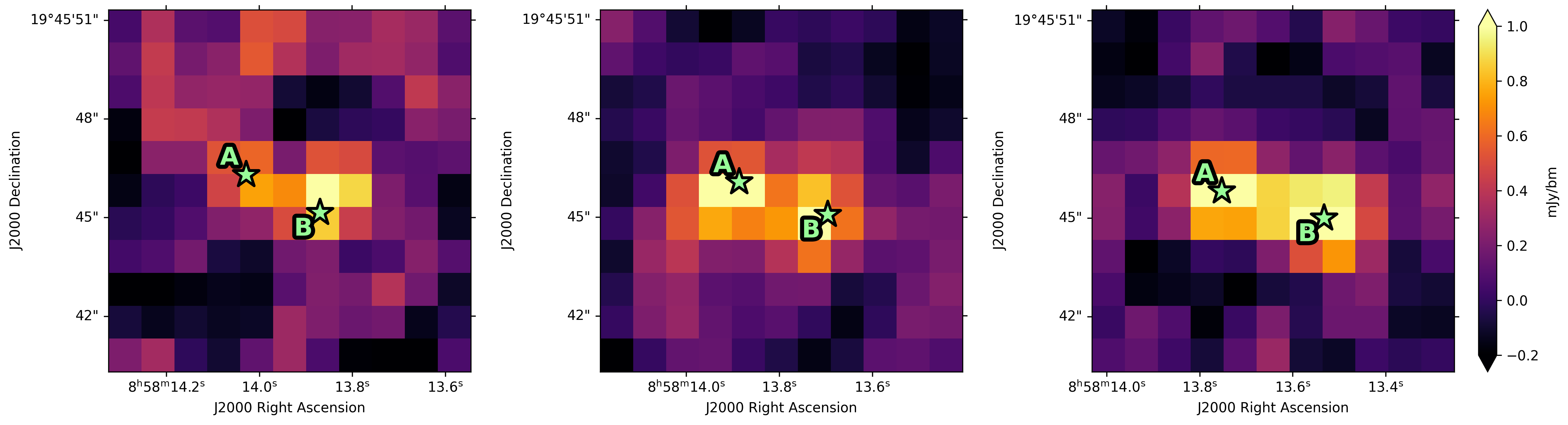} 
    \caption{Detections of \EI A and B during VLASS epochs 1, 2, and 3, i.e., in 2019, 2021, and 2024 (left to right). The positions of the stellar components are labeled, corrected for proper motion. Both stars are detected in all three epochs. See Table~\ref{table:ArchivalRadio} for associated measurements and calculations. }
\label{fig:VLASS}
\end{center}
\end{figure}

As displayed in Table~\ref{table:ArchivalRadio}, \EIAB has been detected at higher frequencies over the past three decades, with observations from 1993 to 2024 spanning 855.5 to 5000 MHz and extending into the microwave regime.
Within the 31 years of published archival radio data, only resolved observations by the VLA Sky Survey (VLASS) reveal emission from both stars (see Figure~\ref{fig:VLASS}). VLASS epoch 3 flux measurements have not yet been released; we extracted flux densities from the publicly available Quick Look image, adopting Canadian Initiative for Radio Astronomy Data Analysis (CIRADA)\footnote{\url{https://cirada.ca/}} \texttt{PyBDSF} parameters to be consistent with the epoch 1 and 2 catalogs. We also force fit the VLASS 1 image for the A component, which was not reported in the data release. For all measurements, we scaled the peak flux values by 5\% to account for flux bias\footnote{see \href{https://library.nrao.edu/public/memos/vla/vlass/VLASS_022.pdf}{VLASS Project Memo \#22}} and added 5\% flux scale uncertainties to the error in quadrature \citep{Perley2017}. Peak fluxes were adopted as flux density to account for fitting biases introduced by the presence of the companion. Emission is expected to a occur at a radius of $\sim 1 R_*$; brightness temperatures for the 3 epochs are $0.7\times 10^{13}-1 \times 10^{13}$ K, approaching the boundary delineating coherent and incoherent emission. Further discussion of the nature of the higher frequency emission can be found in Section~\ref{ssec:mechanisms}.

\section{Discussion} \label{sec:discussion}

The VLITE detection of emission from \EIAB represents the first confirmed detection of a UCD or UCD binary at $\sim$340 MHz. Our results extend the known radio activity of \EIAB to lower frequencies and suggest coherent emission in line with earlier M dwarfs observed at this frequency. However, incoherent emission throughout the radio spectrum cannot be ruled out without additional data. The resolved detection of both stars in our VLASS analysis also reveals that magnetic activity is characteristic of each star rather than of one or the other. Here we discuss (1) possible mechanisms of the emission and (2) stellar rotation measurements that are unusual for the magnetic and fundamental properties of the system.

\subsection{Possible Emission Mechanisms} \label{ssec:mechanisms}

\subsubsection{Incoherent Emission: Gyrosynchrotron Radiation}
\label{sssec:gyro}

Although the stellar disk-averaged brightness temperatures for our VLITE observations suggest coherent emission, it is possible that the emission events occurred on an extended magnetic loop and correspond to incoherent emission (see Section~\ref{ssec:eicancri_vlite}).

The G\"{u}del-Benz empirical relation (GBR) \citep{Drake1989,GuedelBenz1993,BenzGuedel1994} is a nearly linear relation between X-ray luminosity and radio luminosity, which can be tied to accelerated electrons in magnetic reconnection events such as stellar flares. Incoherent gyrosynchrotron emission is generally associated with the relation \citep[e.g.,][]{Vedantham2022} and is common in magnetically active M dwarfs \citep{Callingham2021}. At EI~Cancri's X-ray luminosity of 9.2647$\times10^{27}$ erg~s$^{-1}$, the GBR predicts a gyrosynchrotron radio luminosity of approximately $3\times10^{12}$ erg~s$^{-1}$~Hz$^{-1}$ corresponding to about 0.1 mJy for \EI AB. Although \EIAB is a highly magnetically active system with frequent flares (see Section~\ref{ssec:mag_rot}), our VLITE measurements of 2.70 mJy - 14.4 mJy are radio-loud compared to the GBR. This deviation from the GBR is in line with other findings of active M dwarfs \citep[][140 MHz]{Pope2021, Callingham2021} and UCDs \citep[][5-9 GHz]{Berger2001, Williams2014} below the relation, and has been predicted for coherent auroral emission in UCDs \citep{Hallinan2008}.

This implication of coherent emission would appear to rule out the possibility that the emission occurs at a larger loop, which would correspond to a brightness temperature indicative of incoherent emission. However, although deviation from the GBR strongly suggests coherent emission, incoherent emission can occur without adherence to the GBR \citep[e.g.,][]{Williams2014}. Furthermore, our VLITE observations are not resolved between the two stellar components, and one star may be more radio quiet than the other. Follow-up observations at 340 MHz to measure properties such as polarization would further clarify the nature of the system's emission at this frequency (see Section~\ref{sec:conclusion}).

At higher frequencies, \EIAB appears to exhibit quiescent emission, with detections tracing back as far as 1993 (Table~\ref{table:ArchivalRadio}). Higher frequency emission originates near the stellar disk of the star and its small scale magnetic loops, but the unresolved nature of the majority of the detections makes it unclear what proportion of the flux corresponds to each star. 
The exceptions to this limitation are the detections of each star in the VLASS 3 GHz observations. 
With flux densities ranging from 0.82 mJy - 1.706 mJy, the emission detected by VLASS is more radio loud than the GBR prediction of 0.1 mJy, suggesting coherent emission from each star.

In summary, with the available data, we lack strong indication of incoherent gyrosynchrotron, but cannot rule it out either in VLITE or higher frequency data. Coherent emission across a broad range of frequencies as implied by deviation from the GBR would be consistent with previous M dwarf findings by \cite{VilladsenHallinan2019}. However, it is possible that incoherent emission could be present without adherence to the GBR. In such a magnetically active binary, a combination of both coherent and incoherent emission is possible across the radio spectrum, and further simultaneous multi-frequency observations, with polarization, are needed to evaluate the breakdown between these two types of emission.

\subsubsection{Coherent Emission: ECMI vs. Plasma Emission}

A coherent emission mechanism such as ECMI or plasma emission at 340 MHz is suggested by the stellar-disk-averaged brightness temperature and deviation of \EIAB from the GBR (Sections~\ref{ssec:eicancri_vlite} \& \ref{sssec:gyro}). Distinguishing between ECMI and plasma emission is frequently done by analyzing the periodicity of the signal, how broadband the emission is, the frequency drift, and/or polarization properties of the emission \citep[e.g.,][]{Lynch2017,VilladsenHallinan2019,Zic2019,Rose2023,Bloot2024,Callingham2024}. We are unable to identify a periodic signal using VLITE with only one epoch of detection. The VLITE band is also only $\sim$40~MHz wide; we detect emission in all three of its sub-bands for the combined 7-hour image. The band is not wide enough to place additional constraints on how broadband the emission is or to check for frequency drift, especially not without potentially diluting the signal below detectability in the time chopped images. VLITE has some polarization capability, but complexity in the VLITE system has hindered development of the analysis process; many studies use the circular polarization fraction to discern between emission mechanisms, but VLITE exhibits leakage between components and non-polarized light. Because VLITE is commensal with other VLA programs, the VLITE program has no control over calibrator observations, making it yet more difficult to disentangle the circular ($V$) and two linear ($Q$ \& $U$) polarization components. It is possible to extract a total polarization value, $P = \sqrt{Q^2+U^2+V^2}$; refinement of the process is underway to be presented in future work.

Although, we cannot directly distinguish between ECMI or plasma emission at this time using VLITE, we suspect that the emission is from ECMI because radio-emitting UCDs have been shown to frequently exhibit coherent emission via the ECMI mechanism \citep{Hallinan2008,VilladsenHallinan2019}  and \EIAB emission is radio loud relative to the GBR (see Section~\ref{sssec:gyro}). Analysis of periodicity, broadband nature, frequency drift, or polarization properties at higher frequencies are beyond the scope of this project, but may be possible in some cases with access to raw datasets. 

\subsection{Magnetic Activity and Rotation} 
\label{ssec:mag_rot}

Radio emission from \EIAB has been detected throughout the past three decades at frequencies ranging from 340 MHz to 220 GHz (see Table~\ref{table:ArchivalRadio}). With all other radio data unresolved between the two stars, resolved VLASS detections provide the first evidence that both \EI A and \EI B are magnetically active. In the following discussion, we expand upon this result with evidence of magnetic activity across the electromagnetic spectrum and highlight stellar rotation measurements that are at odds with these findings.  

\EI has a strong history of magnetic signatures and was first identified as a flare star system by \cite{Pettersen1985}, who detected 24 flaring events within 4.5 hours at 320-400 nm. \cite{Newton2017} also classify this system as ``active'' with an H$_\alpha$ equivalent width (EW$_{H\alpha}$) of $-5.719$. Signs of magnetic activity have also been found in all-sky surveys, exhibited at higher energies in the form of near- and far-ultraviolet emission \cite[Galaxy Explorer; GALEX;][]{Martin2005_GALEX,Bianchi2017_GALEX} and X-ray emission \citep[ROentgen SATellite; ROSAT;][]{Boller2016_ROSAT_2RXS}. These are complemented by frequent optical flaring and starspot modulation present in four sectors of TESS data \citep{10.17909/fwdt-2x66}; although there are some  simultaneous observations with VLITE, we detect no 340~MHz emission during these epochs. The full suite of these activity markers combine with VLITE and archival radio detections to indicate a highly magnetically active system, with resolved VLASS emission the only firm indicator that both stars are active. 

Based on the high magnetic activity in the \EI system, we could expect \EI A and B to be rapid rotators \citep[e.g.,][]{Skumanich1972,Wright2011,Newton2017,Wright2018}. \cite{Jeffers2018}\footnote{referencing Rodr\`{i}guez, H. 2014, Master's thesis, UCM Madrid} report a projected rotational velocity (vsini) of 16.70$\pm$0.64 km~s$^{-1}$ for the system, which could correspond to a period upper limit of 10.76 hours for \EI A or 9.60 hours for \EI B. In contrast, a period of 83.270 days is measured by \cite{Newton2016}, and we calculate rotation periods of 113 days and 116 days, respectively, for \EI A and B using the \cite{Newton2017} mass-rotation relation. 

This is notable because a 10-hour rotation period is compatible with expectations for a magnetically active M dwarf, while a rotational period of 83 days would be strikingly long. This is evidenced in the \cite{Newton2017} mass-rotation-activity diagram (Fig. 5), where a $\sim10$-hour period implies an active star, while an 83-day period suggests more likely inactivity (although the sample in that area of phase space is small). Similarly, we can estimate an X-ray to bolometric luminosity ratio for this system of $L_X/L_{bol} \sim 1.5\times10^{-3}$ \citep[adopting log L$_{bol}\sim-2.8$ based on][]{Pecaut2012,MamajekTable}\footnote{\url{https://github.com/emamajek/SpectralType/blob/master/EEM_dwarf_UBVIJHK_colors_Teff.txt}} and find that it is consistent with the majority of fully convective stars at the shorter $\sim10$-hour period, versus uncharacteristically high at an 83-day period  \citep[see Figure 2 of][]{Wright2018}. 

It is unclear what each period can be ascribed to in this system; it is possible that the two different timescales each correspond to a rotation period, with one star at the more-expected $\sim$10 hours and the other at 83 days. This would be an unusual mismatch between two otherwise similar stars in the same system and at odds with the detection of magnetic activity in both stars, as resolved by VLASS. However, discrepant rotation periods for more identically twin M dwarfs have been measured before by \cite{Couperus2025,Couperus2026}, although the largest mismatch is 6.55 days and 38 days. In twin systems, mismatches in rotation and activity indicators can be driven by multiple factors, including stellar dynamo stochasticity or star-planet interaction. To characterize the possible discrepancy between observed magnetic activity and stellar rotation rate in the \EI system, we recommend resolved observations to measure rotation periods for each star.

\subsection{Summary}

We have discovered UCD emission at 340 MHz for the first time with the VLITE detection of \EI AB. Although we suspect coherent ECMI emission as its origin, additional data are needed to discern between coherent and incoherent emission and to rule out plasma emission and gyrosynchrotron radiation at this frequency. Detected radio emission at all frequencies is radio-loud relative to the GBR, implying coherent emission, although it is possible for incoherent emitters to deviate from the GBR. Furthermore, the unresolved flux measurements may be distributed unevenly between the stars; future resolved observations may place one star or the other on the GBR and imply gyrosynchrotron emission, while the other is radio-loud and more likely to be coherent ECMI. The VLASS detections are the only dataset with individual flux densities for each star, revealing that both stars are radio emitters and magnetically active. Although the brightness temperature measurements lie at the boundary and cannot be used as clear indicators for coherence versus incoherence, both stars are radio loud compared to the GBR, suggesting coherent emission. Coherent ECMI emission across a broad range of frequencies aligns with previous findings for M dwarfs, while gyrosynchrotron emission would also align with the frequent flares exhibited by the system. 

Adding a layer of intrigue to \EI AB, the system has a visible light photometric signal with an 83-day period and a projected rotational velocity measurement corresponding to up to about 10 hours; while a 10-hour rotation period aligns with other active M dwarf findings, the case of an 83-day period for either star would be anomalous given proof from VLASS that both stars are magnetically active.  

\section{Conclusions and Future Work} \label{sec:conclusion}

We identify radio emission at 340 MHz from the \EI system using VLITE, marking the first detection of an UCD at this frequency. To date, $\sim$300-MHz frequency studies have detected only a handful of earlier-type M dwarfs. Since 1993, the \EI system has been detected at higher frequencies, ranging from 855.5 MHz to 5.0 GHz and including microwave at 90-220 GHz. Expanding the detection frequency down to 340 MHz for the first time reveals that the ionized electrons may extend further out from the stellar surface, suggesting a larger-scale magnetic field. Although we cannot disentangle which star is the emitter for each event, each event is point-like and is unlikely to be emission from both stars. Each of the three events could be either coherent or incoherent, depending on the radius of emission, which could be at the stellar disk or on a magnetic loop extending as far as $10~ R_*$ or further. Based on findings at MHz frequencies for earlier M dwarfs, the emission is most likely coherent and auroral in nature, produced via the ECMI mechanism, although we cannot presently rule out plasma emission or gyrosynchrotron radiation. This system is highly magnetically active, and incoherent emission is likely tied to flaring events. \EI also emits at 3 GHz, with resolved emission from both stars detected by VLASS confirming that both \EI A and B are radio emitters. 

To further characterize the nature of \EI at 340 MHz will require additional data with higher sensitivity and broader wavelength and/or time coverage to track frequency drift and/or search for a periodic signal. One option is pointed observations using the VLA WIDAR system at P-band; centered at a similar frequency, the VLA WIDAR P-band system provides a wider bandpass than VLITE and therefore is more sensitive to fainter emission. Also, pointed, on-axis observations would not suffer the attenuation we correct for in our VLITE observations (0.874$^\circ$ from pointing center). In addition, without the effects of off-axis polarization leakage, we could properly calibrate Stokes parameters. Lastly, we would have access to the full 27 antennas in the array, rather than the 18 antennas used for VLITE. This would yield more collecting area and uv coverage. Summarily, pointed VLA observations in P-band could lead to higher S/N, polarization information, and improved time resolution of the rise and decay of events at $\sim$300 MHz \citep[see for example,][]{VilladsenHallinan2019}. Another option is to take pointed GHz frequency observations using the VLA to be combined with simultaneous VLITE observations at 340 MHz. Refinement of the VLITE polarization data processing procedures could also yield a measurement of the polarization fraction for the detections reported in this paper to discern between emission mechanisms.  Branching out to other facilities, Very Long Baseline Array\footnote{\href{https://science.nrao.edu/facilities/vlba}{https://science.nrao.edu/facilities/vlba}} observations could be acquired to map out the orbit of the system, while also probing $\sim300$~MHz, with options ranging from P-band to 96 GHz. A spectrum using archival \EIAB detections is challenging to interpret due to uncertainty in the fraction of emission coming from each component and in the time evolution of the emission. Resolved radio observations spanning a range of frequencies would facilitate the construction of a spectrum for each star. In this highly magnetically active system with multiple possible rotation periods, it would also be valuable to determine the rotation period of each star and identify whether they are uncharacteristically long (see Section~\ref{ssec:mag_rot}). This could be done by measuring projected rotational velocity via high resolution infrared spectroscopy observations with a 1\arcsec or smaller slit. Another compelling case is to observe the system using Keck adaptive optics with the infrared wavefront sensor. These data would provide resolved infrared flux measurements on each star and constrain the parameter space for any additional UCD components in the system. Critically, they could be used in understanding their full spectral properties and derive fundamental parameters such as effective temperature, bolometric luminosity, and radius, key to putting the star in context with the greater UCD population.

\begin{acknowledgements}
    We thank Dr.~Rachel Osten for insightful discussions on UCD magnetic activity at radio frequencies. We also thank our referee for their comments, which contributed to the interpretation of our results and improved the clarity of their presentation. 
    
    This scientific work uses data obtained from Inyarrimanha Ilgari Bundara / the Murchison Radio-astronomy Observatory. We acknowledge the Wajarri Yamaji People as the Traditional Owners and native title holders of the Observatory site. CSIRO’s ASKAP radio telescope is part of the Australia Telescope National Facility (https://ror.org/05qajvd42). Operation of ASKAP is funded by the Australian Government with support from the National Collaborative Research Infrastructure Strategy. ASKAP uses the resources of the Pawsey Supercomputing Research Centre. Establishment of ASKAP, Inyarrimanha Ilgari Bundara, the CSIRO Murchison Radio-astronomy Observatory and the Pawsey Supercomputing Research Centre are initiatives of the Australian Government, with support from the Government of Western Australia and the Science and Industry Endowment Fund. This paper includes archived data obtained through the CSIRO ASKAP Science Data Archive, CASDA (https://data.csiro.au).

    The National Radio Astronomy Observatory is a facility of the National Science Foundation operated under cooperative agreement by Associated Universities, Inc. CIRADA is funded by a grant from the Canada Foundation for Innovation 2017 Innovation Fund (Project 35999), as well as by the Provinces of Ontario, British Columbia, Alberta, Manitoba and Quebec.
    
    This work has made use of data from the European Space Agency (ESA) mission {\it Gaia} (\url{https://www.cosmos.esa.int/gaia}), processed by the {\it Gaia} Data Processing and Analysis Consortium (DPAC, \url{https://www.cosmos.esa.int/web/gaia/dpac/consortium}). Funding for the DPAC has been provided by national institutions, in particular the institutions participating in the {\it Gaia} Multilateral Agreement.

    This paper includes data collected with the TESS mission, obtained from the MAST data archive at the Space Telescope Science Institute (STScI). Funding for the TESS mission is provided by the NASA Explorer Program. STScI is operated by the Association of Universities for Research in Astronomy, Inc., under NASA contract NAS 5–26555.

    This project was supported in part by an appointment to the NRC Research Associateship Program at the U.S. Naval Research Laboratory in Washington, D.C., administered by the Fellowships Office of the National Academies of Sciences, Engineering, and Medicine. 

    Basic research at NRL is funded by 6.1 Base programs. Construction and installation of VLITE was supported by the NRL Sustainment Restoration and Maintenance fund. 

\end{acknowledgements}

\facilities{VLITE, VLA, TESS, GALEX, ROSAT, Gaia, ACT, ASKAP}

\software{Python, \texttt{NumPy} \citep{NumPy2020}, \texttt{Matplotlib} \citep{Matplotlib2007}, \texttt{Astropy} \citep{Astropy2022}, \texttt{Obit} \citep{Cotton2008_Obit}, \texttt{AIPS} \citep{Greisen2003_AIPS}, \texttt{PyBDSF} \citep{Mohan2015}}

\bibliography{references}{}
\bibliographystyle{aasjournal}

\end{document}